\documentclass[aps,pra,superscriptaddress,showpacs,twocolumn,footinbib]{revtex4-1}
\usepackage{amsmath,amsfonts,amssymb}
\usepackage{graphicx,float,calc}
\usepackage{color,bm}
\usepackage{ulem}
\usepackage[colorlinks,urlcolor=blue,citecolor=blue,linkcolor=blue]{hyperref}

\begin{document}

\author{Zhen-Fei Zheng}
\affiliation{Key Laboratory of Quantum Information, and Synergetic Innovation Center of Quantum Information and Quantum Physics, 
University of Science and Technology of China, Hefei, Anhui 230026, People's Republic of China}

\author{Guang-Can Guo}
\affiliation{Key Laboratory of Quantum Information, and Synergetic Innovation Center of Quantum Information and Quantum Physics,
University of Science and Technology of China, Hefei, Anhui 230026, People's Republic of China}

\author{Han Pu} \email{hpu@rice.edu}
\affiliation{Department of Physics and Astronomy and Rice Center for Quantum Materials, Rice University, Houston, Texas 77251-1892, USA}
\affiliation{Center for Cold Atom Physics, Chinese Academy of Sciences, Wuhan 430071, People's Republic of China}

\author{Xu-Bo Zou} \email{xbz@ustc.edu.cn}
\affiliation{Key Laboratory of Quantum Information, and Synergetic Innovation Center of Quantum Information and Quantum Physics,
University of Science and Technology of China, Hefei, Anhui 230026, People's Republic of China}

\title{Field-induced topological pair-density wave states in a multilayer optical lattice}

\begin{abstract}
We study the superfluid phases of a Fermi gas in a multilayer optical lattice system in the presence of out-of-plane
Zeeman field, as well as spin-orbit (SO) coupling. We show that the Zeeman field combined with the SO coupling leads to exotic topological pair-density wave (PDW) phases in which different layers possess different superfluid order parameters, even though each layer experiences the same Zeeman field and the SO coupling. We elucidate the mechanism of the emerging PDW phases, and characterize their topological properties by calculating the associated Chern numbers.

\end{abstract}

\pacs{67.85.-d, 03.75.Ss, 74.20.Fg}

\maketitle

\section{Introduction}

Ultracold atoms in optical lattices offer a remarkable platform for investigating quantum many-body problems
and simulating solid state materials \cite{simulation1}. The high degree of controllability and tunability of the system
parameters and free of lattice vibrations and structural defects make the optical lattices ideal as analog quantum simulators
\cite{coldatom1,coldatom2,coldatom3,coldatom4,coldatom5}. The optical lattices in the experiments are typically constructed
by interfering several laser beams to realize a fully controllable lattice geometry and the lattice depth is tunable by the laser
intensity. In addition, the tunneling rate between lattice sites can be precisely tailored by microwave pulses or radio-frequency 
fields \cite{laser-hop-rev-1, laser-hop-rev-2,laser-hop-exp-1,laser-hop-exp-2,rf-hop-exp-1}, to realize various exotic lattice models. Furthermore, by employing 
external fields \cite{driven-lattice-rev-1,driven-lattice-rev-2,driven-lattice-rev-3,driven-lattice-scheme-1,driven-lattice-scheme-2,
driven-lattice-scheme-3}, synthetic gauge potentials can be generated. The versatility of driving schemes might enable one to 
explore unconventional phases that are hard to reach in static solid-state systems. Such unconventional phases include the Fulde-Ferrell-Larkin-Ovchinnikov (FFLO) 
phase \cite{fflo-1st-ff,fflo-1st-lo,fflo-ref-1,fflo-ref-2, fflo-ref-3} and the pair-density wave (PDW) phase \cite{pdw-rev,pdw-ori-1,pdw-ori-2,pdw-ori-3}
that have attracted tremendous interest in the past decades.

The PDW state, a novel superfluid state with layer-dependent order
parameters, has been extensively studied in the context of unconventional superconductors \cite{pdw-1,pdw-3,pdw-2,pdw-4,pdw-5} and
is believed to exist in the Ce-based heavy-fermion superconductor such as $\mathrm{CeCoIn}_5$ \cite{ce-1,ce-2,ce-3,ce-4,ce-5}.
It also plays a key role in the formation of color superconductivity in high-density quark matter \cite{ref-qm}. Moreover, 
in the past few years, the topological properties of PDW are widely investigated in both time reversal-breaking and time
reversal invariant systems \cite{pdw-1,ref-topo-1,ref-topo-3,ref-topo-4}, and the systems are classified into $\mathbb{Z}$ and $\mathbb{Z}_2$
topological classes. The topological nature of the PDW phase are protected by symmetries. For example, the PDW state investigated in Ref.~\cite{pdw-1} 
is protected by mirror symmetry in a tri-layer system, and a pair of non-trivial chiral edge
excitations emerge as long as the protecting symmetries are not broken. Thus the PDW states provide an ideal experimental candidate
in the search of symmetry-protected topological phases for interacting fermions. So far such states have not been unambiguously observed in experiment.

Most of the previous theoretical proposals on realizing PDW phases in multi-layer systems are based on a layer-dependent spin-orbit (SO) coupling \cite{pdw-ori-3,pdw-1,pdw-2,ref-topo-4}, which results in a layer-dependent order parameter, i.e., the PDW phase. In cold-atom experiments, the SO coupling is induced by Raman coupling between hyperfine ground states of the atom
\cite{soc-heat-1,soc-heat-2}. However, realization layer-dependent SO coupling remains experimentally challenging. Furthermore, such a scheme may require more laser beams which can cause severe heating to the quantum gases. 
As such, an interesting question that can be raised is the following: Can PDW phases emerge in a multi-layer system with identical SO coupling across all layers?

In this paper, we address this question and show that indeed topological PDW phases can emerge in ultracold Fermi gases in multi-layered lattice systems with layer-independent SO coupling, together with an out-of-plane (i.e., perpendicular to the layers) Zeeman field. 
The paper is organized as follows.
We present the model Hamiltonian in Sec. \ref{sec-model} for a bilayer system.
In Sec. \ref{sec-phase} we present our numerical results based on the 
self-consistent Bogoliubov-de Gennes (BdG) equation. We discuss the phase diagram
and characterize various phases.
By tuning the Zeeman field, we show how the superfluid order parameter
acquires a spontaneous layer-modulated phase due to inter-band pairing.
In Sec. \ref{sec-topo}, we show that the transition from the BCS to the PDW states is associated with a topological quantum phase transition.
We extend the same study to a tri-layer system in Sec. \ref{sec-trilayer}. Finally, Sec. \ref{sec-conclude} is devoted to the conclusions
and some final remarks.

\section{The model}\label{sec-model}

The physical system we consider here is a uniform SO coupled degenerate spin-1/2
Fermi gas confined in a two-dimensional (2D)
bilayer square optical lattice with an out-of-plane Zeeman field. In the tight-binding limit, the system can be
described by the following Fermi-Hubbard Hamiltonian
\begin{equation}
  H = H_0 + H_{so} + H_I, \label{eq-real-H}
\end{equation}
where the single-particle Hamiltonian $H_0$ reads
\begin{align}
H_0 = &-\mu\sum_{i,m} \hat{n}_{i\sigma m} -t\sum_{\langle ij\rangle,m} \hat{c}^{\dag}_{i\sigma m}
\hat{c}_{j\sigma m} \notag\\
&+ h_z\sum_{i,m}\left(\hat{n}_{i\uparrow m}-\hat{n}_{i\downarrow m}\right)
- t_{\perp}\sum_{i\sigma} \left( \hat{c}^{\dag}_{i\sigma 1}\hat{c}_{i\sigma 2}+h.c. \right),
\end{align}
where $i$ and $j$ label the sites on each layer, $m$ (=1, 2) denote the two layers, $\sigma$ ($= \uparrow$, $\downarrow$) denote the atomic spin states,
$\hat{c}_{i\sigma m}$ is the particle annihilation operator
 at site $i$ with spin $\sigma$ on layer $m$, and $\hat{n}_{i\sigma m}$ is the particle number
 operator. In Hamiltonian (\ref{eq-real-H}), $t$, $\mu$, $h_z$ and $t_{\perp}$ are the intra-layer tunneling amplitude, the chemical potential, the out-of-plane
Zeeman field strength and the inter-layer hopping strength, respectively. The two tunneling amplitudes $t$ and $t_\perp$ are taken to be non-negative. The SO coupling Hamiltonian takes the Rashba form
 \begin{equation}
 H_{so} = -\frac{\alpha}{2i} \sum_{\langle ij\rangle,m} \hat{\psi}^{\dag}_{im}\left(
 \mathbf{d}_{ij} \times \hat{\sigma}\cdot\mathbf{e}_z\right) \hat{\psi}_{jm},
 \end{equation}
 which couples the spin-up and spin-down components of neighboring sites within each layer with a layer-independent coupling strength $\alpha$.
Here $\mathbf{d}_{ij}$ is the unit vector between site $i$ and $j$, $\mathbf{e}_z$ is
 unit vector along $z$-axis which is perpendicular the layer plane, $\hat{\sigma}$ are the spin Pauli matrices, and $\hat{\psi}_{im}=\left(
 \hat{c}_{i\uparrow m},\hat{c}_{i\downarrow m}\right)^T$.  Such types of the SO coupling and the effective Zeeman field have been theoretically proposed and realized in recent experiments for both bosons and fermions \cite{fflo-ref-2,Rashba-ref-1,Rashba-ref-2, Rashba-ref-3,Rashba-ref-4,Rashba-ref-5}.
Finally, the two-body interaction Hamiltonian takes the form 
 \begin{equation}
 H_I = U\sum_{i,m}\hat{n}_{i\uparrow m}\hat{n}_{i\downarrow m},
 \end{equation}
 where $U<0$ is the attractive interaction strength.
 
 In order to investigate the superfluid phase of the interacting Fermi gas, we
 take the mean-field approximation, in which the two-body
 interaction Hamiltonian can be reduced into
 \begin{equation}
 U\hat{n}_{i\uparrow m}\hat{n}_{i\downarrow m} = \Delta_m\hat{c}^{\dag}_{i\uparrow m}
 \hat{c}^{\dag}_{i\downarrow m} + \Delta^*_m \hat{c}_{i\downarrow m}\hat{c}_{i\uparrow m}
 -|\Delta_m|/U
 \end{equation}
 where $\Delta_m = U\langle \hat{c}_{i\downarrow m}\hat{c}_{i\uparrow m} \rangle$ is 
 the superfluid order parameter for layer $m$.
 
Transforming the mean-field Hamiltonian into momentum space, we have
\begin{align}
H = &\sum_{\mathbf{k},m} \hat{\psi}^{\dag}_{\mathbf{k} m}\left[ \xi(\mathbf{k}) I+\alpha\mathbf{g(k)}
 \cdot \hat{\sigma}+h_z\sigma_z\right]
 \hat{\psi}_{\mathbf{k} m} \notag\\
 &+ \left[ -t_{\perp}\sum_{\mathbf{k},\sigma}
  \hat{c}^{\dag}_{\mathbf{k}\sigma 1}\hat{c}_{\mathbf{k}\sigma 2} 
  +\sum_{\mathbf{k},m} \Delta_m \hat{c}^{\dag}_{\mathbf{k}\uparrow m}
  \hat{c}^{\dag}_{-\mathbf{k}\downarrow m} +h.c. \right]~,
  \label{eq-H-mom}
 \end{align}
 where $\xi(\mathbf{k}) = -\mu -2t\cos{k_x}-2t\cos{k_y}$ is the single-particle dispersion and $\mathbf{g(k)} = 
 (-\sin{k_y},\sin{k_x},0)$. Momentum $\mathbf{k}$ spans the first Brillouin zone with $k_{x,y}\in[-\pi/a,\pi/a]$ and we set
 the lattice constant $a=1$. The momentum-space Hamiltonian can be rewritten as
 \begin{equation}
 H = \frac{1}{2}\sum_{\mathbf{k}} \hat{\Psi}^{\dag}(\mathbf{k})\mathcal{H}(\mathbf{k})
 \hat{\Psi}(\mathbf{k}) - \sum_m\frac{\Delta_m^2}{g} + \sum_{\mathbf{k}}2\xi(\mathbf{k}),
 \label{eq-mf-m}
 \end{equation}
 under the Nambu spinor basis 
 \begin{equation}
 \Psi(\mathbf{k})=(\hat{c}_{k\uparrow 1},\hat{c}_{k\downarrow 1},
 \hat{c}_{k\uparrow 2},\hat{c}_{k\downarrow 2},\hat{c}^{\dag}_{-k\uparrow 1},
 \hat{c}^{\dag}_{-k\downarrow 1}, \hat{c}^{\dag}_{-k\uparrow 2},\hat{c}^{\dag}_{-k\downarrow 2})^{T}
 \end{equation}
 and the Bogoliubov-de Gennes (BdG) operator
 \begin{equation}
 \mathcal{H}(\mathbf{k}) = 
\left( \begin{array}{cc}
 \hat{\mathcal{H}}_0(\mathbf{k}) & \hat{\Delta}(\mathbf{k})\\
 \hat{\Delta}^{\dag}(\mathbf{k}) & -\hat{\mathcal{H}}^{T}_0(-\mathbf{k})
 \end{array} \right), \label{eq-bdg-matrix}
 \end{equation}
 where 
\begin{eqnarray*}
\mathcal{H}_0(\mathbf{k}) &=&  (\xi(\mathbf{k}) I+\alpha\mathbf{g(k)}
 \cdot \hat{\sigma}+h\sigma_z)\otimes \hat{\tau}_I-t_{\perp}I\otimes \hat{\tau}_x\,,\\
\Delta(\mathbf{k}) &=& i\Delta_m \hat{\sigma}_y \otimes \hat{\tau}_I\,,
\end{eqnarray*}
with $\hat{\tau}$ being the Pauli matrices acting on the layer space. As usual,
 the mean-field Hamiltonian (\ref{eq-mf-m}) can be diagonalized by 
 the BdG transformation \[\hat{c}_{\sigma m} (\mathbf{k})= 
 \sum_{\sigma \eta}(u^\eta_{\sigma m}(\mathbf{k})
 \gamma_{\eta \sigma m}+v_{\sigma m}^{\eta}(\mathbf{k})\gamma^{\dag}_{\eta \sigma m}),\]
 with quasiparticle operators $\gamma_{\eta \sigma m}$ and $\gamma^{\dag}_{\eta \sigma m}$.
 The BdG quasiparticle spectrum are obtained by diagonalizing $\mathcal{H}(\mathbf{k})$:
 \begin{equation}
 \mathcal{H}(\mathbf{k})\phi_{\eta}(\mathbf{k}) = E_{\eta} \phi_{\eta}(\mathbf{k}),\label{eq-bdg}
 \end{equation}
 with quasiparticle energies $E_{\eta}$ and wave functions $
 \phi_{\eta}(\mathbf{k})=[u^{\eta}_{\uparrow 1},u^{\eta}_{\downarrow 1},
 u^{\eta}_{\uparrow 2},u^{\eta}_{\downarrow 2},v^{\eta}_{\uparrow 1},
 v^{\eta}_{\downarrow 1},v^{\eta}_{\uparrow 2},
 v^{\eta}_{\downarrow 2}]^{T}.$
 
 We numerically solve Eq.~(\ref{eq-bdg}), and self-consistently determine $\Delta_m$. When several solutions are obtained,
 the ground state is determined by the one that renders the lowest energy. We characterize the phases by the values of $\Delta_m$.
 When ${\Delta_{1}=\Delta_2=0}$, the system is in a non-superfluid normal gas (NG) state. When $\Delta_1
 = \Delta_2 \neq 0$, the system is in a BCS state. When $\Delta_1 = -\Delta_2 \neq 0$,
 the system is in a PDW state.

\section{Phase diagram}\label{sec-phase}

 \begin{figure}[tbp]
\centering
\includegraphics[width=0.5\textwidth]{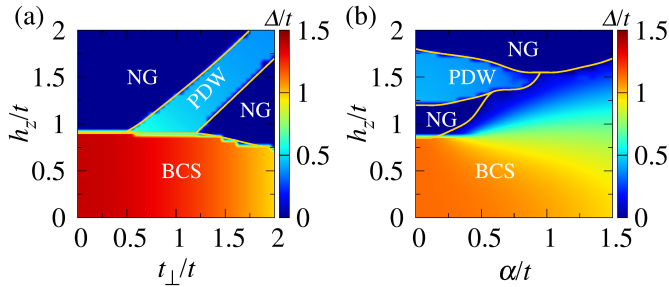}
\caption{(Color online) 
(a) Phase diagram of the bilayer system in the $t_{\perp}$-$h_z$ plane by setting the SO coupling strength $\alpha = 0$. (b) Phase diagram of the bilayer system in the $\alpha$-$h_z$ plane by setting the inter-layer hopping strength $t_{\perp} = 1.5t$.
Other parameters are $\mu = 0$ and $U = -4t$. The order parameter for the BCS and the PDW phases are given by $(\Delta, \Delta)$ and $(\Delta, -\Delta)$, respectively.}
\label{fig-phase}
\end{figure}

In Fig. \ref{fig-phase} we present the zero temperature ground state phase diagram. Let us first consider the case with $\alpha=0$, i.e., in the absence of SO coupling. The phase diagram in the $t_\perp$-$h_z$ plane is shown in Fig.~\ref{fig-phase}(a) where we fix the interaction strength to be $U=-4t$ and the chemical potential to be $\mu=0$ which means the system is in half-filling. Furthermore, the system is spin balanced when $h_z = 0$ and spin imbalanced when $h_z \neq 0$. 
When the Zeeman field is very small, the system favors the normal BCS superfluids.
With the increase of the Zeeman field $h_z$, the BCS state becomes unstable and the system either becomes normal (NG) or enters the PDW phase. The existence of the PDW phase requires a finite inter-layer hopping $t_\perp$ that is comparable to $h_z$.

\begin{figure}[tbp]
\centering
\includegraphics[width=0.45\textwidth]{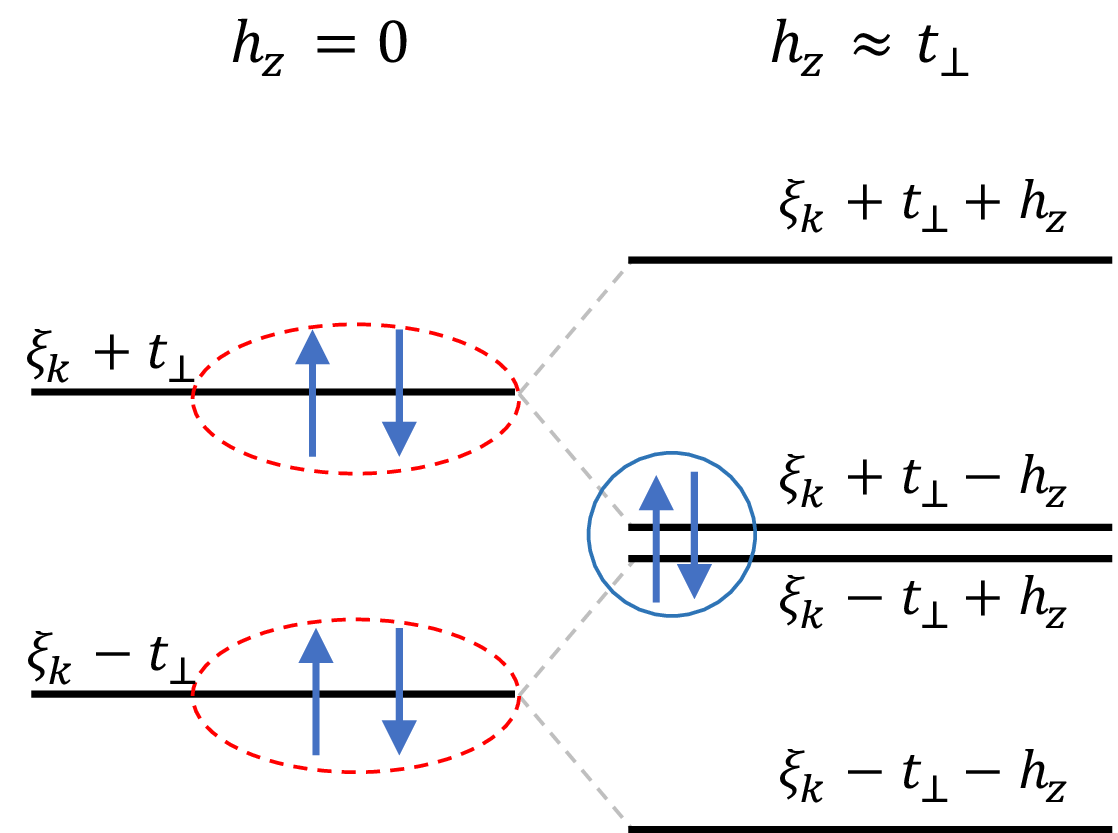}
\caption{(Color online) 
The single particle dispersion in the absence of the SO coupling. Without the Zeeman field (left panel), pairing between opposite spins occur within the same energy band, as shown by the red dashed circles. When $h_z$ is comparable to $t_\perp$ (right panel), the inter-band pairing (shown by the blue solid circle) becomes favored.} 
\label{fig-pairing}
\end{figure}

The mechanism for the emergence of the PDW phase can be understood as follows. In the absence of the SO coupling, the single-particle dispersion can be easily obtained as
\begin{equation}
E_{\mathbf{k}}^{a,b} = \xi(\mathbf{k}) \pm t_{\perp} + \sigma h_z,
\label{eq-spec}
\end{equation}
where $a$ and $b$ labels the two bands due to the hybridization of the two layers via inter-layer hopping, $\sigma = \pm$ for spin-up and spin-down atoms. The corresponding particle creation operators are given by 
\begin{align}
\label{bilayer}
\psi^{\dag}_{a,\sigma}(\mathbf{k}) &= (c^{\dag}_{\mathbf{k}\sigma,1}+c^{\dag}_{\mathbf{k}\sigma,2}) /\sqrt{2},\notag\\
\psi^{\dag}_{b,\sigma}(\mathbf{k}) &= (c^{\dag}_{\mathbf{k}\sigma,1}-c^{\dag}_{\mathbf{k}\sigma,2})/\sqrt{2}.
\end{align}
For zero Zeeman field $h_z=0$, each band is two-fold degenerate due to the degeneracy of the spin-up and spin-down atoms (see the left panel of Fig.~\ref{fig-pairing}). In this case, superfluid pairing occurs within each band, as schematically represented by the red dashed circles in Fig.~\ref{fig-pairing} and the system is in the usual BCS phase. As $h_z$ increases, the energies of the spin-up and spin-down atoms start to split (the former shifted up and the latter shifted down), which eventually destabilizes the superfluid pairing within each band
\footnote{In principle, the Zeeman field could induce an FFLO phase, see, for example, A. Buzdin, S. Tollis, and J. Cayssol, Phys. Rev. Lett. {\bf 95}, 167003 (2005). However, the parameter regime where the FFLO phase is stable is very small, hence we ignore it in the present work.}. However, when $h_z$ becomes comparable to $t_\perp$, the spin-down atoms from the top band and the spin-up atoms from the bottom band become nearly degenerate. This gives rise to an inter-band pairing, as illustrated by the blue solid circle in Fig.~\ref{fig-pairing}. As seen from Eq.~(\ref{bilayer}), there exists a relative $\pi$ phase difference between layer 1 and layer 2. As a result, the inter-band pairing leads to the PDW phase where the order parameters in the two layers possess opposite signs.  

Next, we consider the effect of the SO coupling. In Fig.~\ref{fig-phase}(b), we fix $t_\perp=1.5t$, and present the phase diagram in the $\alpha$-$h_z$ parameter space. As the SO coupling term mixes opposite spins, it enhance the intra-band pairing and reduce the inter-band pairing. The general effect of the SO coupling is to favor BCS pairing. Consequently, as $\alpha$ increases, the BCS regime expands and the PDW regime shrinks. However, Fig.~\ref{fig-phase}(b) fails to capture one crucial effect of the SO coupling, that is it can drive the superfluid phase into a topological one. We now turn to a more detailed discussion of the topological property of the system.

\section{Topological property}\label{sec-topo}

Let us first define the time-reversal
operator $\mathcal{T} = i \hat{\sigma}_y\mathcal{K}$, the particle-hole operator $\mathcal{P} = \hat{s}_x \mathcal{K}$, and the chiral symmetry operator
$\mathcal{C} = \mathcal{P}\mathcal{T}$ with $\mathcal{K}$ the complex conjugation operator and $\hat{\mathbf{s}}$ the Pauli matrices acting on the particle-hole
space. One could find that the total Hamiltonian in the presences of the SO coupling preserves the particle-hole symmetry, while breaking the time-reversal symmetry and the chiral symmetry.
According to the generic classification scheme \cite{topo-class}, the system belongs to the D class in 2D
and can be characterized by Chern number $\mathcal{C}$ \cite{chern-def-1,chern-def-2,chern-def-3,chern-def-4}.
Due to the degeneracy of energy bands at some points in Brillouin zone, the Chern number is non-Abelian and is defined by
\begin{equation}
\mathcal{C}_{\psi} = \frac{1}{2\pi}\int_S d^2\mathbf{k}~\mathrm{Tr}~\mathrm{d}\mathcal{A}, \label{eq-chern}
\end{equation}
with gap opening-condition. Here $\mathrm{d}\mathcal{A}$ is defined as $\mathrm{d}\mathcal{A} = \partial_{k_x}\mathcal{A}_{k_y} - \partial_{k_y}\mathcal{A}_{k_x}$, where the non-Abelian Berry connection is given by $\mathcal{A}_{\mu} = -i\langle \psi|\partial_{\mu}|\psi
\rangle$ ($\mu = k_x, k_y$) which is an $M\times M$ matrix, with $|\psi \rangle = (|\phi_i\rangle,...,|\phi_{i+M}\rangle)^T$ representing a vector of eigenvectors of $M$ occupied bands (typically, $M = 2\sim 4$ in our calculation depending on the specific parameters).

\begin{figure}[tbp]
\centering
\includegraphics[width=0.50\textwidth]{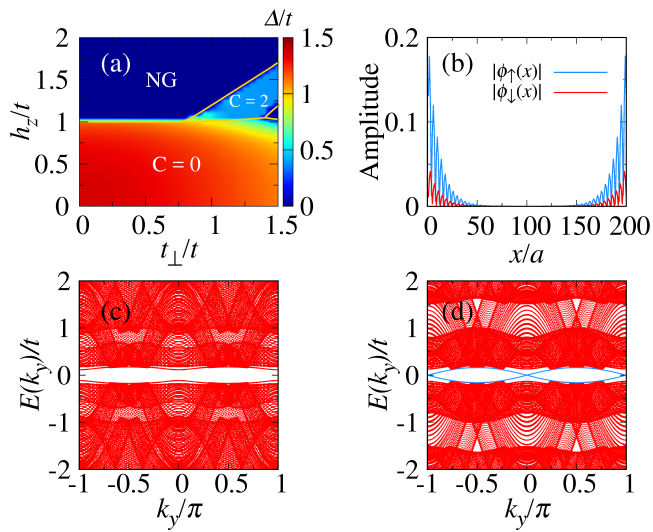}
\caption{(Color online) 
(a) Phase diagram in $h_z$-$t_{\perp}$ plane at fixed SO coupling strength $\alpha = 0.5t$, chemical
potential $\mu = 0$ and interaction strength $U = -4t$. The color scheme represents
the magnitude of the order parameter $|\Delta_{1,2}|=\Delta$. (b) Wave function of the topological edge states
localized near the two edges of the system supported by the topological PDW state whose spectrum is shown in (d). The amplitude of the spin-resolved wave function
in the two layers are the same. (c) The spectrum of the non-topological BCS state at $t_\perp=1.5t$ and $h_z=0.5t$. 
(d) The spectrum of the topological PDW state at $t_\perp=1.5t$ and $h_z=1.5t$. The blue
solid lines in the gap represent a pair of chiral edge states. In (b), (c) and (d), a hard-wall potential is added in the $x$-axis.} 
\label{fig-topo}
\end{figure}

In Fig. \ref{fig-topo}(a) we plot the phase diagram in the $h_z$-$t_{\perp}$ plane with fixed SO coupling $\alpha = 0.5t$ at chemical
potential $\mu = 0$ and interaction strength $U = -4t$. We find that the BCS phase in this region is topologically trivial and the
PDW phase is topologically nontrivial with Chern number $\mathcal{C} = 2$. The topology of the superfluid state can be determined by examining the excitation gap $\Gamma$,
which is defined by $\Gamma = \mathrm{min}\{|E_{\eta}| \}$, where $E_{\eta}$ is the quasiparticle energy defined in Eq.~(\ref{eq-bdg}). It describes the gap between the particle and the hole bands.
For the BCS state, the gap $\Gamma$ closes at $h_z = \sqrt{\Delta^2+t_{\perp}^2}$, and system is non-topological when $h_z < \sqrt{\Delta^2+t_{\perp}^2}$ with $\mathcal{C} = 0$ and topological when $h_z > \sqrt{\Delta^2+t_{\perp}^2}$ with $\mathcal{C} = 4$. Our calculation shows that for the parameters represented in Fig. \ref{fig-topo}(a), we always have $h_z < \sqrt{\Delta^2+t_{\perp}^2}$ and hence a non-topological BCS phase. For the PDW state, the gap closes at two critical Zeeman field strengths $h_1=-\Delta + t_\perp$ and $h_2=\Delta + t_\perp$. The PDW state is non-topological if $h_z<h_1$ with $\mathcal{C} = 0$, and topological if $h_1 < h_z < h_2$ and $h_z>h_2$ with $\mathcal{C} = 2$ and 4, respectively. For the parameters given in Fig.~\ref{fig-topo}(a), we find that $h_1 < h_z < h_2$ is satisfied for the PDW phase. 

In order to verify the bulk-edge correspondence, we add a hard-wall potential along the $x$-axis and solve the BdG equation to find the quasiparticle spectrum. Two representative spectra are shown in Fig.~\ref{fig-topo}(c) and (d). The former is the spectrum of a non-topological BCS state, and the latter that of a topological PDW state. The blue solid lines inside the bulk gap in Fig.~\ref{fig-topo}(d) represent a pair of topological edge states. The wave function of the edge state is shown in Fig.~\ref{fig-topo}(b), from which we can see that it is indeed localized near the two edges defined by the hard-wall potential along the $x$-axis.

\section{tri-layer lattice system}\label{sec-trilayer}

\begin{figure}[tbp]
\centering
\includegraphics[width=0.50\textwidth]{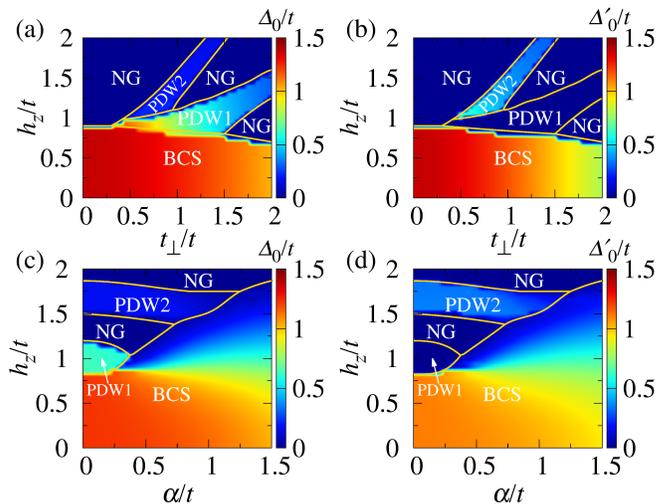}
\caption{(Color online)
(a)-(b) Phase diagram for a tri-layer optical lattice system in the $t_{\perp}$-$h_z$ with the SO coupling strength
$\alpha = 0$. The color describes the amplitude of (a) $\Delta_0$ and (b) $\Delta'_0$. (c)-(d) Phase diagram in the $\alpha$-$h_z$ plane at the fixed inter-layer hopping strength $t_{\perp} = 1.2t$. The color describes the magnitude of (c) $\Delta_0$ and (d) $\Delta'_0$.
Other parameters are $\mu = 0$ and $U = -4.0t$. The order parameter for the BCS, the PDW$_1$, and the PDW$_2$ phases are given by $(\Delta_0, \Delta'_0, \Delta_0)$, $(\Delta_0, 0, -\Delta_0)$, and $(\Delta_0, -\Delta'_0, \Delta_0)$, respectively.}
\label{fig-triphase}
\end{figure}

We now extend the same analysis to a tri-layer square optical lattice system. 
The Hamiltonian of the tri-layer system has the same form as in the previous case, shown in Eq. \ref{eq-real-H}, with the exception that the layer index $m$ now takes values 1, 2, and 3. The middle layer 2 is coupled to the top and the bottom layers (1 and 3, respectively) by the inter-layer hopping, while the top and the bottom layers are not coupled directly between themselves. Again, the phase can be referred from the values of the order parameters obtained from the self-consistent BdG equation. The normal phase (NG) is characterized by $\Delta_1=\Delta_2=\Delta_3=0$, the BCS phase by ($\Delta_1,\Delta_2,\Delta_3$) = ($\Delta_0,\Delta'_0,\Delta_0$), where $\Delta_0$ and $\Delta'_0$ have the same phase but different magnitude. We identify two PDW phases which we denote as $\mathrm{PDW}_1$ and $\mathrm{PDW}_2$. In $\mathrm{PDW}_1$, we have 
($\Delta_1,\Delta_2,\Delta_3$) = ($\Delta_0,0,-\Delta_0$); and in $\mathrm{PDW}_2$, ($\Delta_1,\Delta_2,\Delta_3$) 
= ($\Delta_0,-\Delta_0',\Delta_0$). The difference between the BCS phase and the $\mathrm{PDW}_2$ phase is that, in the former, the order parameter does not change sign when one goes from one layer to the next; whereas, in the latter, it does change sign.

In Fig. \ref{fig-triphase}(a)-(b), we present the phase diagram of the tri-layer
system in the $t_\perp$-$h_z$ plane in the absence of the SO coupling. Similar to the previous case, the BCS phase occupies the regime with small $h_z$. For large $h_z$, the BCS state is unstable and the system is either in the normal phase or one of the two PDW phases. 

\begin{figure}[tbp]
\centering
\includegraphics[width=0.49\textwidth]{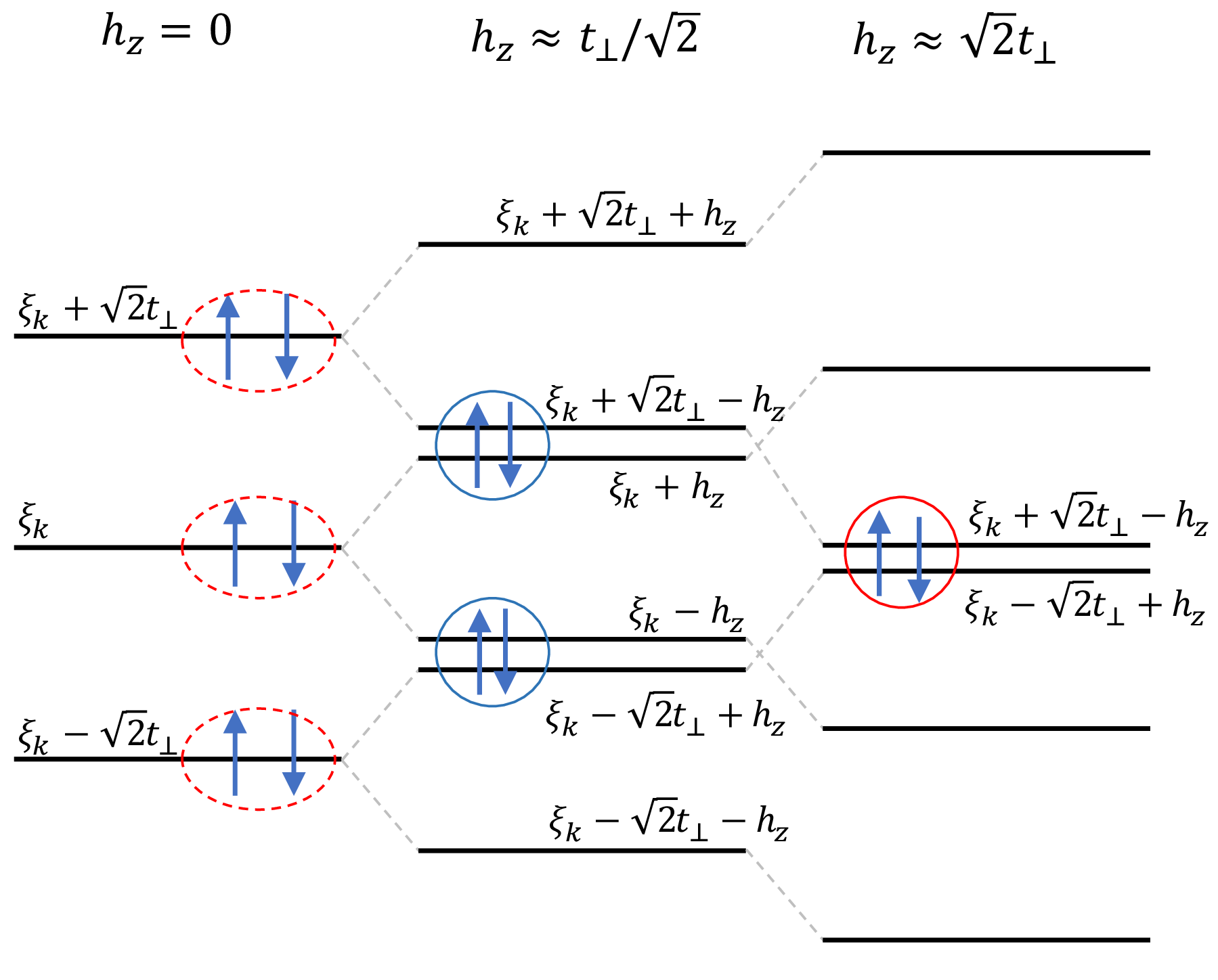}
\caption{(Color online) 
The single particle dispersion of the tri-layer system in the absence of the SO coupling. Without the Zeeman field (left panel),
the pairing between opposite spins occur within the same band (red dashed circles). When the Zeeman field strength is increased such that $h_z \approx t_\perp/\sqrt{2}$, the pairing between two adjacent bands is favored (blue solid circle), which leads to the $\mathrm{PDW}_1$ phase. Further increase the Zeeman field such that $h_z \approx \sqrt{2} t_\perp$, the pairing is favored between the top and the bottom band (red solid circle), and $\mathrm{PDW}_2$ phase emerges.} 
\label{fig-me-tri}
\end{figure}

The emergence of the two PDW phases can be understood in a similar way as in the previous case. 
In the absence of the SO coupling,
the single particle dispersion in the three bands (labeled as $a$, $b$ and $c$) take the form
\begin{equation}
E_{\mathbf{k}\sigma}^{a,c}=\xi(\mathbf{k}) \pm \sqrt{2}t_{\perp} +\sigma h_z, \;\;\;E_{\mathbf{k}\sigma}^{b} = \xi(\mathbf{k}) +\sigma h_z,
\end{equation}
with the corresponding creation operators given by 
\begin{align}
\psi^{\dag}_{a,\sigma} &= (c^{\dag}_{\mathbf{k}\sigma,1}+\sqrt{2}c^{\dag}_{\mathbf{k}\sigma,2}+c^{\dag}_{\mathbf{k}\sigma,3})/2,\notag\\
\psi^{\dag}_{b,\sigma} &= (c^{\dag}_{\mathbf{k}\sigma,1}-c^{\dag}_{\mathbf{k}\sigma,3})/\sqrt{2},\notag\\
\psi^{\dag}_{c,\sigma} &= (c^{\dag}_{\mathbf{k}\sigma,1}-\sqrt{2}c^{\dag}_{\mathbf{k}\sigma,2}+c^{\dag}_{\mathbf{k}\sigma,3})/2.
\end{align}
At zero (or small) $h_z$, as represented by the left panel of Fig.~\ref{fig-me-tri}, pairing mainly occurs within the same band and the conventional BCS phase is realized. Increase the Zeeman field such that $h_z \approx t_\perp/\sqrt{2}$ (middle panel of Fig.~\ref{fig-me-tri}), inter-band pairing between the top and the middle bands and that between the middle and the bottom bands become resonant and hence favored. The lack of $c_{\mathbf{k}\sigma,2}^\dag$ component in $\psi_{b,\sigma}^\dag$ and the relative $\pi$ phase difference between the $c_{\mathbf{k}\sigma,3}^\dag$ component in $\psi_{b,\sigma}^\dag$ and $\psi_{a(c),\sigma}^\dag$ explains the emergence of the $\mathrm{PDW}_1$ phase with ($\Delta_1,\Delta_2,\Delta_3$) = ($\Delta_0,0,-\Delta_0$). Alternatively, one can interpret the absence of the order parameter in the middle layer 2 as due to the destructive interference via its coupling to the other two layers 1 and 3. Finally, further increase $h_z$ to near $\sqrt{2}t_\perp$ favors the inter-band pairing between the top and the bottom bands (right panel of Fig.~\ref{fig-me-tri}). The relative $\pi$ phase difference between the $c_{\mathbf{k}\sigma,2}^\dag$ components in $\psi_{a,\sigma}^\dag$ and $\psi_{c,\sigma}^\dag$ leads to the $\mathrm{PDW}_2$ phase with ($\Delta_1,\Delta_2,\Delta_3$) = ($\Delta_0,-\Delta'_0, \Delta_0$).

\begin{figure}[tbp]
\centering
\includegraphics[width=0.45\textwidth]{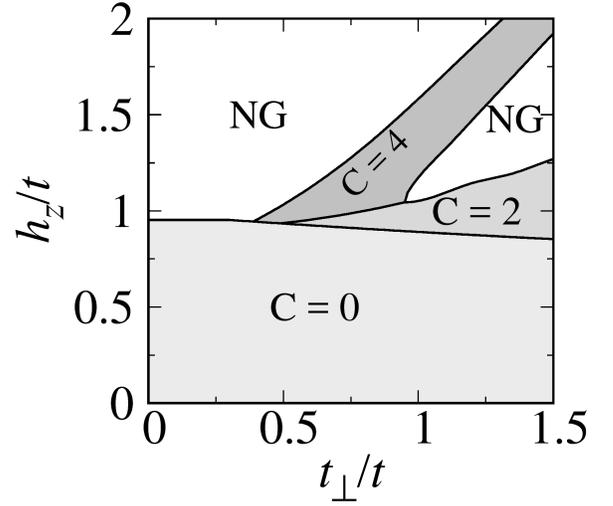}
\caption{(Color online) Phase diagram in the $t_{\perp}$-$h_z$ plane at fixed SO coupling $\alpha = 0.25t$.  The Chern numbers for various superfluid phases are indicated.
The Chern numbers for BCS phase is $0$, $2$ for $\mathrm{PDW}_1$ phase and $4$ for $\mathrm{PDW}_2$ phase.
The other parameters are $\mu = 0$ and $U = -4t$.} 
\label{fig-chern-tri}
\end{figure}

Also similar to the previous bilayer system, the presence of the SO coupling enhances BCS pairing and suppress the PDW phases, as shown in Fig.~\ref{fig-triphase}(c)-(d), and the resulting PDW phases are topological, belonging to class D characterized by nonzero Chern numbers. This is shown in Fig.~\ref{fig-chern-tri}. The two PDW phases possess different Chern numbers: $\mathcal{C} = 2$ for $\mathrm{PDW}_1$, and 4 for $\mathrm{PDW}_2$. Therefore, transitions among different superfluid phases represented in  Fig.~\ref{fig-chern-tri} are also topological phase transitions.

\section{conclusion}\label{sec-conclude}

In conclusion, we have considered both a bi- and a tri-layer square optical lattice system of spin-1/2 Fermi gas subject to an out-of-plane
Zeeman field and Rashba SO coupling. Both the Zeeman field and the SO coupling strengths are uniform across different layers. Nevertheless, PDW phases with layer-dependent order parameter can be realized. Furthermore, the interplay between the Zeeman field and the SO coupling gives rise to topological PDW phases, which can be characterized by finite Chern numbers. We stress again that our proposal is very different from those proposed in Refs.~\cite{pdw-1,pdw-2} where a layer-dependent SO coupling produces
the PDW superfluid. Our scheme, which utilizes a layer-independent SO coupling, has the advantage of simplicity. Our work can have important implications 
in the search of symmetry-protected topological PDW states in multi-layer systems.

\section{Acknowledgements}

We would like to thank Bin Wang and Zhen Zheng for helpful discussions.
This work is supported by National Natural Science Foundation of China (Grants No. 11674305 and No. 11474271),
National Key R$\&$D Program (Grants No. 2016YFA0301300 and No. 2016YFA0301700) and 
Young Scientists Fund of the National Natural Science Foundation of China (Grant No. 11704367).
H. Pu acknowledges support from the US NSF and the Welch Foundation (Grant No. C-1669).

\bibliography{ref}

\end{document}